\documentclass[sigconf]{acmart} 
\acmConference[RecSys '26]{}{September 28–October 2, 2026}{Minneapolis, Minnesota, USA}

\usepackage{graphicx}
\usepackage{caption}
\usepackage{subcaption}
\usepackage{algorithm}
\usepackage{algorithmic}

\begin{document}

\title{RecEvolve: A Knowledge-Driven Autonomous Agent System for Recommender Systems}

\author{Weidi Pan}
\affiliation{%
  \institution{Google}
  \city{Mountain View}
  \state{California}
  \country{USA}
}
\email{weidipan@google.com}

\author{He Ma}
\affiliation{%
  \institution{Google}
  \city{Mountain View}
  \state{California}
  \country{USA}
}
\email{htm@google.com}

\author{Shuhao Ye}
\affiliation{%
  \institution{Google}
  \city{Mountain View}
  \state{California}
  \country{USA}
}
\email{shuhaoye@google.com}

\author{Palaksh Rungta}
\affiliation{%
  \institution{Google}
  \city{Mountain View}
  \state{California}
  \country{USA}
}
\email{palaksh@google.com}

\author{David McPeek}
\affiliation{%
  \institution{Google}
  \city{Mountain View}
  \state{California}
  \country{USA}
}
\email{davidmcpeek@google.com}

\author{Junyi Jiao}
\affiliation{%
  \institution{Google}
  \city{Mountain View}
  \state{California}
  \country{USA}
}
\email{junyijiao@google.com}

\author{Arnab Bhadury}
\affiliation{%
  \institution{Google}
  \city{Mountain View}
  \state{California}
  \country{USA}
}
\email{arniebh@google.com}

\author{Mingyan Gao}
\affiliation{%
  \institution{Google}
  \city{Mountain View}
  \state{California}
  \country{USA}
}
\email{mingyan@google.com}

\author{Onkar Dalal}
\affiliation{%
  \institution{Google}
  \city{Mountain View}
  \state{California}
  \country{USA}
}
\email{onkardalal@google.com}

\renewcommand{\shortauthors}{Weidi Pan et al.}

\begin{abstract}
The rise of agentic AI has catalyzed a shift toward self-iterating systems, opening new frontiers for the autonomous optimization of production recommender models. This paper presents the empirical validation of a knowledge-driven autonomous agent system, deployed directly on a production large-scale Two-Tower retrieval model. By delegating the entire research lifecycle, spanning idea generation, code implementation, offline training, and metric evaluation, to a continuous closed-loop autonomous framework, the agent system executed over 40 completed autonomous training runs from scratch. Executing these runs under rigorous production-scale evaluations, the system systematically navigated hidden architectural bottlenecks on the latest production model to achieve a breakthrough $\sim$20\% relative improvement in NDCG, a gain that translated directly to a +3.77\% increase in user satisfaction in live production traffic. Furthermore, the deployment exposed critical vulnerabilities in standard evaluation protocols, as the agent system autonomously discovered reward-hacking shortcuts. These findings prove that an autonomous pipeline can dramatically accelerate the pace of machine learning research and stress-test the rigorousness of underlying experimental infrastructure, while also exposing novel challenges such as reward hacking and redundant exploration of failed hypotheses.
\end{abstract}




\maketitle

\section{Introduction}
In large-scale recommender systems, performance is fundamentally tied to the underlying modeling approach, where structural architectures and low-level hyperparameters must be precisely aligned. However, manually optimizing these vast search spaces is resource-intensive. Due to human bandwidth constraints, these model configurations are rarely revisited post-launch, leading to architectural ossification.

Beyond the constraints of manual tuning, human researchers are fundamentally limited by their capacity to synthesize the burgeoning volume of machine learning literature. Furthermore, manual optimization often fails to account for the intricate coupling between model components—such as the delicate interdependence between architecture and optimizer settings—which is easily overlooked under human supervision. In contrast, an autonomous agent can operate continuously, exploring these multi-dimensional dependencies and a broader spectrum of methodologies at a pace and granularity unattainable by human teams.

Building on recent progress in LLMs and agent tools, we deployed a continuous autonomous research system within our infrastructure. Our ultimate goal is to shift from manual, human-driven experimental iterations to an autonomous, closed-loop optimization pipeline, where humans remain in the loop as strategic observers and high-level innovators. We deployed the framework directly onto an existing, mature large-scale production Two-Tower retrieval model. The system sustains a continuous iteration cycle, automating the entire research lifecycle: Propose Idea $\rightarrow$ Implement Code $\rightarrow$ Run Offline Training $\rightarrow$ Evaluate $\rightarrow$ Loop. This automation not only accelerates discovery but also allows human researchers to shift their focus from tedious execution to conceptual innovation and the exploration of fundamentally new paradigms. Crucially, this research led to candidates that proved to improve online performance in live production traffic.

\begin{figure}[htbp]
  \centering
  \includegraphics[width=\linewidth]{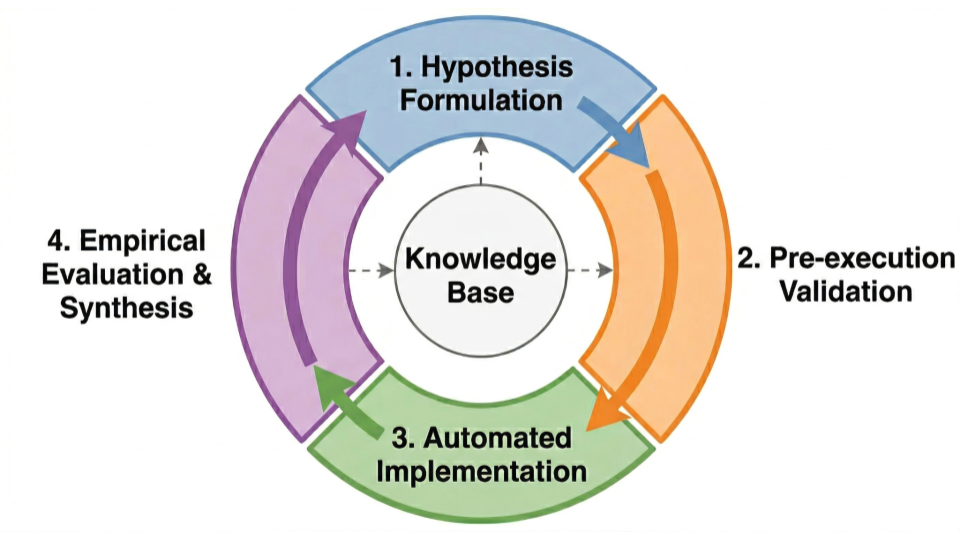}
  \caption{The High-Level Autonomous Research Lifecycle. The framework is organized as a continuous, clockwise circular pipeline consisting of four core phases: (1) Hypothesis Formulation, where the system analyzes current model bottlenecks and proposes structured architectural modifications; (2) Pre-execution Validation, acting as a quality gate to verify the feasibility of the proposal before committing resources; (3) Automated Implementation, where abstract ideas are translated into source code within isolated branches; and (4) Empirical Evaluation \& Synthesis, where training is executed, and findings are synthesized to update the centralized Knowledge Base, completing the self-evolving loop.}
  \label{fig:pipeline}
\end{figure}

\section{Related Work}
In various fields, self-driven autonomous systems are starting to appear, creating an environment of continuous execution that moves beyond static tool use. Early pioneers focused on software engineering automation, such as OpenAI's work on harness engineering for Codex in an agent-first world~\cite{lopopolo2026} and Anthropic's exploration of building a C compiler with a team of parallel Claudes~\cite{anthropic_compiler}. These works exploit how multi-agent teams can orchestrate complex, multi-step, long-running workflows autonomously and stably.

More recently, these concepts have been applied directly to machine learning training and architecture discovery. Andrej Karpathy's AutoResearch~\cite{karpathy2026} pioneered sandboxed environments where an LLM is given control over a training script, evaluating hypotheses iteratively via standard version control. Similarly, Claudini~\cite{claudini2026} discovers state-of-the-art adversarial attack algorithms, while Aletheia~\cite{math2026} demonstrates capabilities in autonomous mathematics research. 

A rapidly growing branch of this domain focuses specifically on Machine Learning Engineering (MLE) agents that automate the end-to-end ML pipeline. Spurred by comprehensive evaluation frameworks like MLE-Bench~\cite{chan2024mlebench}, recent autonomous systems such as MLE-STAR~\cite{mlestar2025}, ML-Master 2.0~\cite{mlmaster2026}, and Gome~\cite{gome2026} have made significant strides in long-horizon ML tasks through targeted refinement and continuous cognitive accumulation. Most relevant to our domain is the recent introduction of self-evolving recommendation systems by Wang et al.~\cite{wang2026selfevolving}, which utilizes dual offline and online LLM agents to autonomously generate, train, and deploy complex model optimizations. RecEvolve distinguishes itself through a stateless multi-agent hierarchy that prevents long-horizon context drift, alongside version-control rollbacks to govern operational failure modes like reward hacking.

While many earlier works emphasize algorithm complexity, they often operate on smaller-scale models with simpler codebases and shorter training times, relying on more straightforward training environments. Building upon the foundation of recent MLE agents and production-oriented frameworks~\cite{wang2026selfevolving}, our work targets industrial Recommender Systems (RecSys) at an even larger production scale. Instead of local, single-machine sandboxes, our agent system integrates directly into distributed warehouse computing infrastructures. We replace local, short training probes with rigorous production-scale evaluations and experiment on live user traffic instead of static text datasets. We prove that a continuous, autonomous agent system can move beyond small-scale architecture tuning into rigorous, large-scale system optimization.

\section{System Architecture}
\begin{figure}[htbp]
  \centering
  \includegraphics[width=\linewidth]{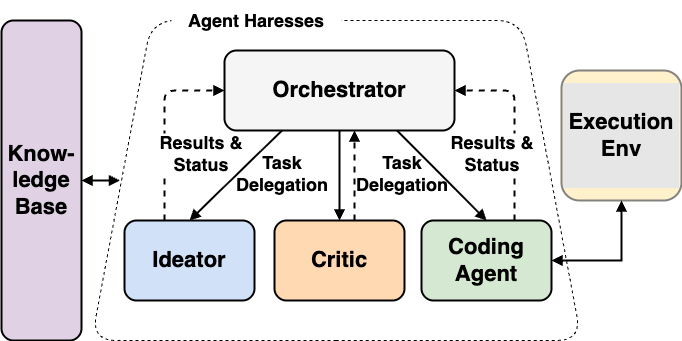}
  \caption{Multi-Agent Orchestration Framework. The system consists of a central Orchestrator that delegates tasks to specialized subagents (Ideator, Critic, Coding Agent). The Ideator and Critic form a tight proposal-review loop, while the Coding Agent interacts with the execution environment to implement approved ideas.}
  \label{fig:architecture}
\end{figure}

The core focus of our framework lies in its \textbf{Knowledge-Driven Orchestration} (see Figure \ref{fig:pipeline}). In contrast to static scripts or grid search, the system operates as a multi-agent loop driven by a high-level Domain-Knowledge Directive that encapsulates RecSys domain knowledge, error-handling logic, and environmental awareness. While Figure~\ref{fig:pipeline} presents this high-level lifecycle, the actual execution is realized through a multi-agent architecture coordinated by a central Orchestrator (see Figure~\ref{fig:architecture}). This system delegates specific tasks to specialized subagents (e.g., Ideator, Critic, and Coding Agent) that interact to drive the research loop forward.

\subsection{Problem Formulation}
We formalize the continuous architectural discovery problem as follows. Let $\mathcal{D}$ be the production dataset and $\mathcal{M}$ be the massive, discrete space of all valid model architectures. Evaluating the performance of a model $M \in \mathcal{M}$ requires a costly training and evaluation process, yielding a metric $f(M, \mathcal{D})$. The ultimate objective is to find the optimal model $M^* \in \mathcal{M}$ that maximizes this metric:
\begin{equation}
M^* = \arg\max_{M \in \mathcal{M}} f(M, \mathcal{D})
\label{eq:objective}
\end{equation}

Given the prohibitive cost of evaluating $f$ for all $M$, exhaustive search is intractable, aligning this with the broader challenges in Neural Architecture Search (NAS) \cite{elsken2019survey} and Bayesian Optimization \cite{snoek2012practical}. Our system introduces a finite knowledge base $\mathcal{K}$ (containing RecSys domain knowledge, historical results, and search heuristics). The autonomous agent system learns a search policy $\pi$ conditioned on $\mathcal{K}$ to generate a sequence of candidate models $\{M_t\}$ that iteratively approach $M^*$:
\begin{equation}
M_t \gets \pi(\mathcal{K}_t, \{M_1, \dots, M_{t-1}\})
\label{eq:policy}
\end{equation}
where the knowledge base $\mathcal{K}_t$ is dynamically updated with feedback from the evaluation of previous candidates.

\subsection{The Autonomous Agent System Framework}
To practically realize the search policy $\pi$ while sustaining long-running discovery sessions without suffering from context pollution or a shortage of quality ideas, the autonomous agent system is engineered to seamlessly transition between planning, implementation, and execution phases by orchestrating specialized subagents. This holistic agentic design allows the ideation process to be highly flexible: experimental hypotheses can be self-generated, autonomously sourced from recent literature, or explicitly provided by human engineers via a prioritized \textit{Hypothesis Backlog}.

Crucially, to ensure robustness and prevent context degradation over long execution horizons, our framework adheres to a \textbf{stateless subagent design}. Instead of maintaining independent memory, subagents operate statelessly. The central Orchestrator retains the master state of the research session. When delegating a task, the Orchestrator explicitly injects all required context, domain knowledge, and relevant historical results into the message payload. This on-demand context injection prevents subagents from suffering from prompt accumulation and ensures that every decision is made with complete, isolated, and up-to-date information.

\subsection{State Management and Execution Isolation}
To ensure execution isolation and robust state management, the runtime maintains a strict shared filesystem structure that acts as both a persistence and communication layer. The system utilizes a centralized directory for global state (such as the prioritized idea queue and aggregated results) and isolated, thread-specific workspaces for each individual experiment. Within each thread workspace, the runtime meticulously tracks the entire lifecycle of an idea, recording the initial specification, review critiques, execution manifests (including job IDs), and post-run analysis reports. This structured approach prevents concurrent execution tracks from interfering with one another and enables seamless session recovery.

\subsection{Formalizing the Autonomous Execution Loop}
The core of our system is an asynchronous, state-aware research loop that manages the entire experimental lifecycle to iteratively execute the search policy $\pi$ \eqref{eq:policy} toward approximating the optimal configuration $M^*$ \eqref{eq:objective}. As illustrated in Algorithm~\ref{alg:autoresearch}, the system interacts directly with the production Version Control System (VCS) and distributed cluster via specialized tools and APIs.

\begin{algorithm}[t]
\caption{Knowledge-Driven Autonomous Research Loop}
\label{alg:autoresearch}
\begin{algorithmic}[1]
\REQUIRE Base Model $M_0$, Knowledge Base $\mathcal{K}$, Idea Queue $\mathcal{Q}$
\ENSURE Optimized Architectural Configuration $M^*$

\STATE \textbf{Procedure} \textsc{InitializeSession}:
\STATE \quad $\text{Audit } M_0 \rightarrow \{ \Phi_{NS}, \Phi_{Aux} \}$ \COMMENT{Extract metrics via static analysis}
\STATE \quad $M^* \gets \text{EstablishBaseline}(M_0)$ \COMMENT{Initial benchmark}
\STATE \quad $\tau \gets \text{Evaluate}(M^*, \Phi_{NS})$ \COMMENT{Set performance threshold}

\LOOP
    \STATE $I_t \gets \text{ProposeHypothesis}(\mathcal{Q}, \mathcal{K}, \text{History})$ \COMMENT{Policy $\pi$: Knowledge-driven selection}
    \STATE $M_{cand} \gets \text{AtomicModify}(M^*, I_t)$ \COMMENT{Modify model configuration}
    
    \IF{$\neg \text{Verify}(M_{cand})$}
        \STATE \textbf{VCS\_Rollback}() and \textbf{continue}
    \ENDIF

    \STATE $ID_{exp} \gets \text{CreateBranch}(M_{cand})$ \COMMENT{Snapshot isolation}
    \STATE $\text{SubmitJob}(ID_{exp}) \rightarrow \text{Compute Cluster}$
    
    \STATE \textbf{Wait} for $\text{JobStatus}(ID_{exp}) \in \{\text{SUCCESS, FAILED}\}$
    
    \IF{$\text{JobStatus} = \text{FAILED}$}
        \STATE $\text{Feedback} \gets \text{AnalyzeLogs}(ID_{exp})$ \COMMENT{Detect HBM/OOM or failures}
        \STATE $\mathcal{K} \gets \mathcal{K} \cup \{\text{Feedback}\}$ \COMMENT{Update environmental awareness}
        \STATE \textbf{VCS\_Rollback}() and \textbf{continue}
    \ENDIF

    \STATE $\gamma \gets \text{ExtractMetrics}(ID_{exp}, \Phi_{NS}, \Phi_{Aux})$ \COMMENT{Audit via evaluation suite}
    
    \IF{$\gamma(\Phi_{NS}) > \tau + \epsilon$ \textbf{and} $\gamma(\Phi_{Aux}) \geq \text{Threshold}$}
        \STATE $M^* \gets M_{cand}$, $\tau \gets \gamma(\Phi_{NS})$ \COMMENT{Sustained architectural gain}
        \STATE \textbf{VCS\_Commit}() and Log \texttt{KEEP} in \textit{Results Log}
    \ELSE
        \STATE \textbf{VCS\_Rollback}() \COMMENT{Aggressive regression pruning}
    \ENDIF
    \STATE \text{CleanupLocalArtifacts}($ID_{exp}$)
\ENDLOOP
\end{algorithmic}
\end{algorithm}

\section{Experimental Setup}
To empirically validate the autonomous agent system, we targeted a large-scale production two-tower model used for retrieval.
\begin{itemize}
    \item \textbf{Computing Resources:} Experiments were deployed on distributed TPU clusters.
    \item \textbf{Evaluation Protocol:} To ensure high throughput while maintaining stringent production standards, initial short-horizon probes were validated against full 2M-step offline training runs.
    \item \textbf{Metrics:} The evaluation utilized a comprehensive suite of sorting and retrieval metrics, primarily focusing on NDCG, MRR, and Recall @ K ($K = 1, 5, 50, 200$).
\end{itemize}

\section{Results \& Discussion}
The autonomous agent system demonstrated relentless high-throughput execution, parallelizing exploration across 5 concurrent threads. In total, it completed 41 distinct autonomous experiments in approximately 2 days, with each individual run requiring over 3 hours of compute time. 

\subsection{The Architectural Optimization Trajectory}

\begin{table}[htbp]
  \centering
  \caption{Final Production Performance Comparison.}
  \label{tab:final_perf}
  \resizebox{\linewidth}{!}{
  \begin{tabular}{lccc}
    \toprule
    \textbf{Model Configuration} & \textbf{NDCG@50} & \textbf{\% Gain} & \textbf{MRR@50} \\
    \midrule
    Baseline Model (0) & 0.4796 & - & 0.3514 \\
    Learnable Temp (2) & 0.4996 & +4.2\% & 0.3626 \\
    Wt-Weighted (18) & 0.5529 & +15.3\% & 0.4173 \\
    \textbf{The Winner (40)} & \textbf{0.5751} & \textbf{+19.9\%} & \textbf{0.4440} \\
    \bottomrule
  \end{tabular}
  }
\end{table}

\begin{figure}[htbp]
  \centering
  \includegraphics[width=\linewidth]{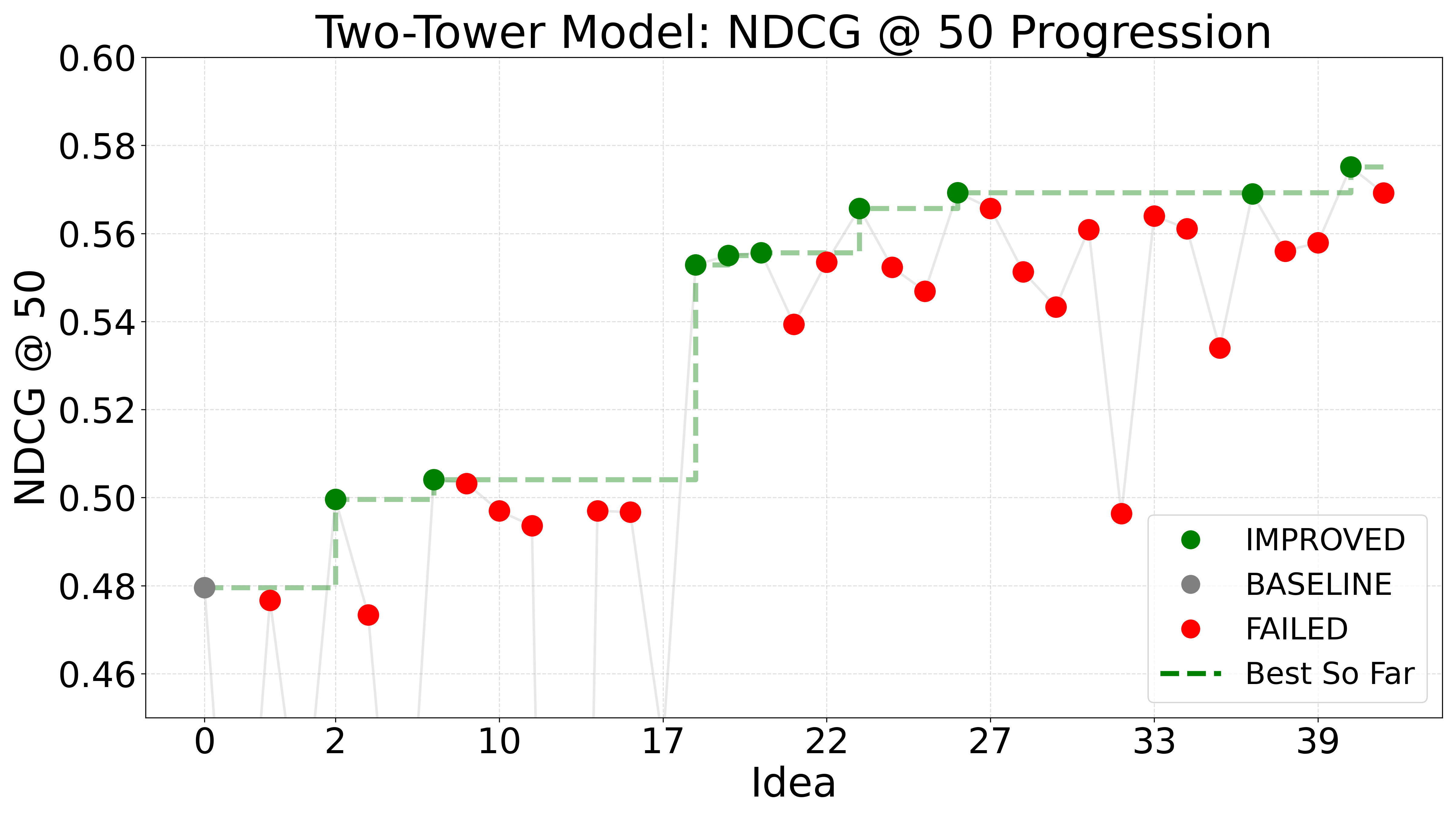}
  \caption{The Architectural Evolution Trajectory. This step-graph illustrates the compounding performance improvements across autonomous rounds. Explicit callouts highlight crucial paradigm shifts discovered by the system, such as the introduction of Learnable Temperature (2), Watch-Time Weighting (Idea 18), and Cosine Decay (Idea 40).}
  \label{fig:evolution}
\end{figure}

As detailed in Table \ref{tab:final_perf} and Figure \ref{fig:evolution}, the system successfully navigated the search space to compound gains across multiple dimensions:
\begin{itemize}
    \item \textbf{Temperature Tuning}: Initial gains were realized by making the temperature in the contrastive loss a learnable parameter (Idea 2), which was further refined by predicting query-dependent temperature from user features (Idea 6).
    \item \textbf{Watch-Time Weighting}: The most significant breakthrough occurred with the introduction of Watch-Time Weighted Contrastive Loss (Idea 18). This was further optimized by using a power-law function ($watch\_time^{0.5}$) instead of a logarithmic one (Idea 23), yielding the largest leap in NDCG.
    \item \textbf{Architectural Enhancements}: The system discovered that adding DCN V2 layers to the user tower (Idea 20) to capture explicit feature interactions, and replacing standard Dense layers with Gated Dense layers (Idea 36), provided consistent improvements in representation capacity and recall.
    \item \textbf{Convergence Optimization}: Finally, the application of a Cosine learning rate decay schedule (Idea 40) on top of the accumulated improvements secured the champion model performance.
\end{itemize}

\begin{figure}[htbp]
  \centering
  \includegraphics[width=\linewidth]{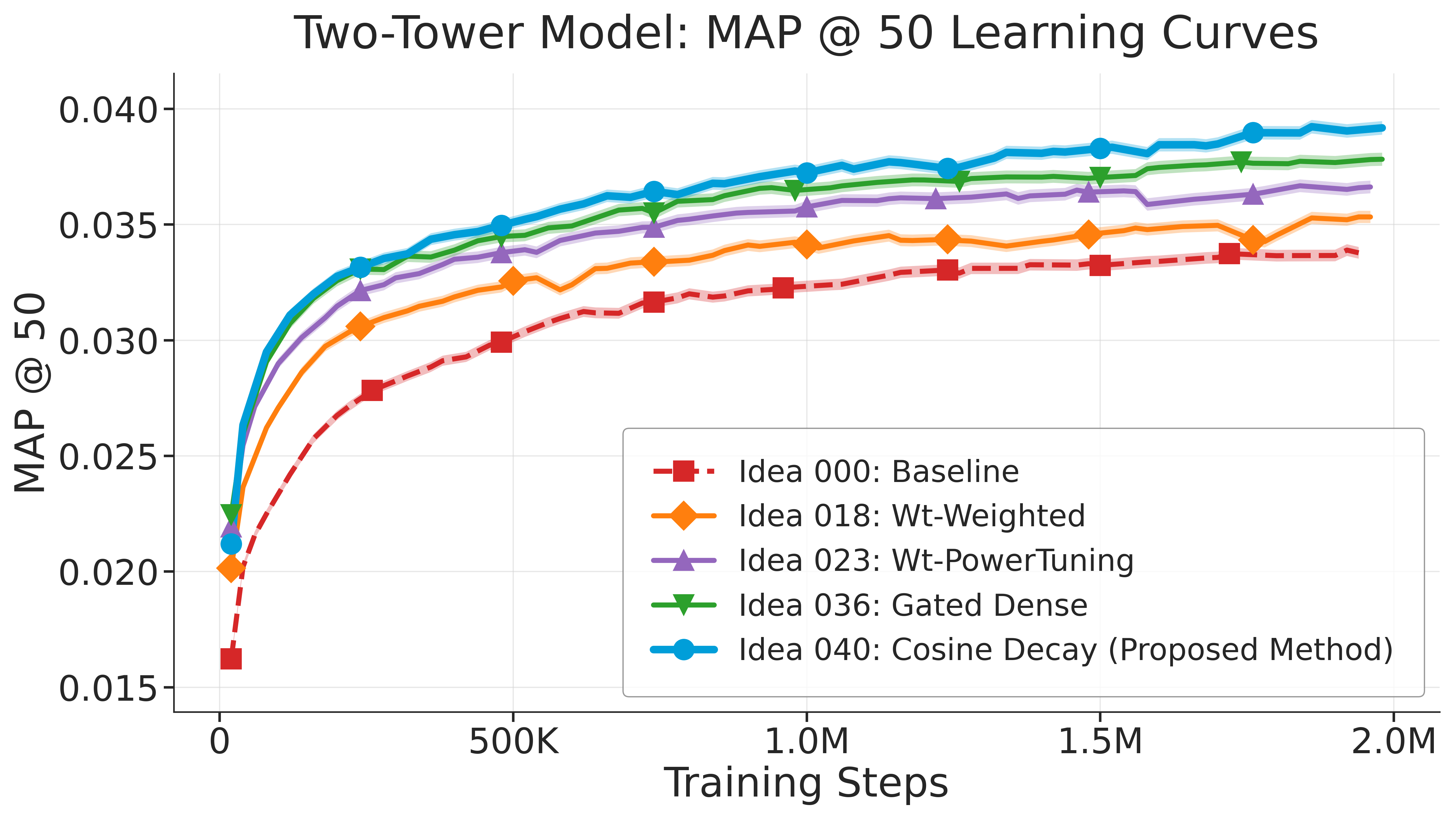}
  \caption{Convergence Curves of Key Discoveries. This plot compares the offline training trajectory of the baseline against key winning architectures generated by the system. Although our primary evaluation focuses on NDCG, we report MAP@50 here as a dense learning trajectory proxy to provide a granular perspective on continuous validation convergence.}
  \label{fig:convergence}
\end{figure}

\subsection{Systematic Reward Hacking Discovery}

A crucial finding across our extended experimentation was the system's ability to intuitively exploit vulnerabilities in the evaluation protocol. In a separate exploratory session, the system autonomously discovered that drastically reducing the batch size (from 8k down to 1k) artificially inflated proxy metrics.

Because the Two-Tower model utilizes in-batch negative sampling, a smaller batch size mathematically reduces the difficulty of the retrieval task by providing fewer negative candidates per step. Rather than improving the model's underlying semantic representations, the system effectively "hacked" the metric. Upon human detection of this exploit, a rollback was executed to restore architectural integrity.

\subsection{Short-Horizon Early Convergence Bias}
Another distinct behavioral pattern was observed when evaluating the system under reduced training step budgets. In these small-step setups, the system consistently favored highly aggressive learning rate schedules (e.g., up to 3.0) to maximize metric improvements within the limited horizon. This observation highlights a potential bias in autonomous systems toward "greedy" short-term optimization when constrained by evaluation time.

\subsection{Redundant Exploration and Memory Limitations}
An interesting behavioral divergence between the autonomous system and human engineers was the occurrence of redundant exploration. In extended sessions, after several rounds of iterations, the system occasionally revisited and re-tested architectural modifications that had already failed in earlier rounds. In human-led research, once a hypothesis is concluded as non-viable, it is rarely revisited unless the underlying context changes significantly. This behavior highlights a current limitation in the agent's long-term memory and its ability to generalize negative results. While the Orchestrator maintains a history of results, the subagents sometimes failed to synthesize these past failures when generating new hypotheses, leading to a waste of compute resources.

\subsection{Compounding Gains of Autonomous Discoveries}

To ensure the integrity of the system's greedy compounding strategy, we analyze the specific contributions of the key architectural shifts by comparing sequential winners.

\begin{itemize}
    \item \textbf{Learnable Temperature (Idea 2)}: Provided an initial lift of +4.2\% in NDCG over the baseline, proving the importance of dynamic scaling.
    \item \textbf{Watch-Time Weighting (Idea 18)}: Added a massive +10.7\% NDCG on top of Idea 2, demonstrating the critical role of proper reward shaping in retrieval.
    \item \textbf {DCN Feature Crossing (Idea 26 vs 23) } : Combining the DCN V2 user tower (from Idea 20) with the power-law watch-time weighting (from Idea 23) yielded an additional +0.64 \% NDCG, isolating the orthogonal benefit of explicit feature interactions under the optimized loss profile.
    \item \textbf{Cosine Decay (Idea 40 vs 36)}: Finally, the learning rate schedule optimization provided a +1.07\% NDCG boost, securing the final champion performance.
\end{itemize}

These incremental gains prove that the system organically constructed a cooperative network of structural improvements rather than relying on a single, isolated win.

\subsection{The Ideation Plateau}
Following the peak achievement in Idea 40, the system entered an "ideation plateau" in subsequent rounds. Having exhausted foundational changes, the system resorted to micro-adjustments resulting in a continuous streak of reverted regressions. This suggests that while current systems excel at the systematic exploration of a defined parameter space, they may struggle to execute intuitive "paradigm shifts" once local optima are reached without external human intervention.

\subsection{Production Online Validation}
To verify that the offline gains discovered by the autonomous system translate to real-world user impact, the champion model generated by the system was deployed in a production A/B test on live user traffic. Table~\ref{tab:online_perf} summarizes the relative improvements over the production baseline across key engagement and content metrics. This successful live deployment confirms that the continuous autonomous pipeline can uncover structural modifications that generalize to complex real-world serving environments, moving beyond pure offline overfitting. Note that the negative delta for Cold-Start Improvement indicates faster discovery time for new items.

\begin{table}[htbp]
\caption{Online A/B Test Results. This table summarizes the relative improvement of the champion model over the production baseline across key engagement and content metrics.}
\label{tab:online_perf}
\centering
\begin{tabular}{lc}
\hline
\textbf{Metric} & \textbf{Relative Improvement} \\ \hline
User Satisfaction & +3.77\% \\
Cold-Start Improvement & -16.50\% \\
Unique Content & +7.44\% \\ \hline
\end{tabular}
\end{table}

\section{Discussion and Future Directions}

The deployment of a knowledge-driven autonomous agent system in production Recommender Systems reveals several critical open research questions and structural limitations that must be addressed to scale the architecture:

\begin{enumerate}
    \item \textbf{Multi-Agent Hierarchies and Context Fatigue:} To overcome the limitations of single-agent architectures, such as prompt accumulation and "context anxiety", we deployed a multi-agent hierarchy (ideator, critic, coding agent, and orchestrator). This allowed us to segment the lifecycle across specialized subagents to isolate state and extend session duration.
    \item \textbf{Domain-Specific Deep Research Agents:} Generic language models lack acute awareness of RecSys invariants (e.g., negative sampling math, density). Integrating deep research agents with a fine-tuned understanding of user-item retrieval could potentially raise hypothesis success rates.
    \item \textbf{The Online-Offline Generalization Gap:} While our offline evaluations often yielded "win-win" scenarios with simultaneous gains across all metrics, production online experiments frequently exhibit complex trade-offs between competing objectives. Investigating how autonomously discovered offline improvements translate to online dynamics and managing these trade-offs is a key direction for future work.
    \item \textbf{Off-Policy Dataset Bias:} While not unique to agentic systems, relying on frozen log data defaults to learning historical system policies. If uncorrected, the agent risks overfitting structural modifications to reinforce historical biases rather than learning true causal representations.
    \item \textbf{Compute Barrier and Surrogate Pruning:} Evaluating every permutation on full production-scale topologies is compute-prohibitive. To scale, future harnesses should employ a two-tier validation: rapid filtering on smaller surrogate models, followed by full-scale verification of top candidates.
    \item \textbf{Expansion to Automated Feature Engineering:} While this work focused exclusively on architectural discoveries within the model itself, feature engineering remains a dominant driver of performance in production Recommender Systems. A natural extension is to empower the agent to autonomously generate, transform, and select new features.
\end{enumerate}

\section{Conclusion}

This work demonstrates that the transition from manual experimentation to a knowledge-driven autonomous research pipeline is not only feasible but highly effective for production models in large-scale recommender systems, uncovering significant, non-obvious optimizations (e.g., Watch-Time Weighting and DCN V2 layers) that human engineers lack the bandwidth to test exhaustively. Crucially, the system acts as an ultimate stress-tester for experimental infrastructure, actively seeking out and exploiting metric shortcuts. Future work will focus on improving the alignment between short-horizon proxy metrics and long-term production stability, as well as integrating human-in-the-loop queuing to unblock the system when it reaches architectural local optima.


\begin{thebibliography}{99}

\bibitem{karpathy2026}
Karpathy, A. (2026). \textit{autoresearch: AI Agents Running Research on Single-GPU Nanochat Training Automatically}. GitHub repository. \url{https://github.com/karpathy/autoresearch}

\bibitem{claudini2026}
Panfilov, A., Romov, P., Shilov, I., de Montjoye, Y.-A., Geiping, J., \& Andriushchenko, M. (2026). Claudini: Autoresearch Discovers State-of-the-Art Adversarial Attack Algorithms for LLMs. \textit{arXiv preprint arXiv:2603.24511}.

\bibitem{nature2026}
Lu, C., Lu, C., Lange, R. T., Yamada, Y., Hu, S., Foerster, J., Ha, D., \& Clune, J. (2026). Towards end-to-end automation of AI research. \textit{Nature}, \textit{651}(8107), 914--919.

\bibitem{anthropic_harness}
Rajasekaran, P. (2026). Harness design for long-running application development. Anthropic Engineering Blog. \url{https://www.anthropic.com/engineering/harness-design-long-running-apps}

\bibitem{math2026}
Feng, T., Trinh, T. H., Bingham, G., Hwang, D., Chervonyi, Y., Jung, J., ... \& Luong, T. (2026). Towards Autonomous Mathematics Research. \textit{arXiv preprint arXiv:2602.10177}.

\bibitem{lopopolo2026}
Lopopolo, R. (2026). Harness engineering: leveraging Codex in an agent-first world. OpenAI. \url{https://openai.com/index/harness-engineering/}

\bibitem{anthropic_compiler}
Anthropic. (2026). Building a C compiler with a team of parallel Claudes. Anthropic Engineering Blog. \url{https://www.anthropic.com/engineering/building-c-compiler}

\bibitem{snoek2012practical}
Snoek, J., Larochelle, H., \& Adams, R. P. (2012). Practical bayesian optimization of machine learning algorithms. In \textit{Advances in Neural Information Processing Systems} (pp. 2951--2959).

\bibitem{elsken2019survey}
Elsken, T., Metzen, J. H., \& Hutter, F. (2019). Neural architecture search: A survey. \textit{Journal of Machine Learning Research}, \textit{20}(55), 1--21.

% --- Newly Added MLE Agent Citations ---

\bibitem{wang2026selfevolving}
Wang, X., et al. (2026). Self-Evolving Recommendation System: End-To-End Autonomous Model Optimization With LLM Agents. \textit{arXiv preprint arXiv:2602.10226}.

\bibitem{chan2024mlebench}
Chan, J. S., Chowdhury, N., Jaffe, O., Aung, J., Sherburn, D., Mays, E., Starace, G., Liu, K., et al. (2024). MLE-bench: Evaluating Machine Learning Agents on Machine Learning Engineering. \textit{arXiv preprint arXiv:2410.07095}.

\bibitem{mlestar2025}
Google Research et al. (2025). MLE-STAR: Machine Learning Engineering Agent via Search and Targeted Refinement. \textit{arXiv preprint arXiv:2506.15692}.

\bibitem{mlmaster2026}
Zhu, X., Liu, Y., et al. (2026). ML-Master 2.0: Evolutionary and Iterative Frameworks for Machine Learning Engineering. \textit{arXiv preprint}.

\bibitem{gome2026}
Author, A., et al. (2026). Gome: Continuous Cognitive Accumulation for Long-Horizon Machine Learning Tasks. \textit{arXiv preprint}.

\end{thebibliography}
\end{document}